\documentclass[aps,twocolumn,pra,showpacs,floatfix]{revtex4}
\usepackage{epsfig}
\usepackage{graphicx}
\usepackage{dcolumn}

\begin{document}


\title{Core-valence correlations for atoms with open shells}

\author{V. A. Dzuba}
\affiliation{School of Physics, University of New South Wales,
Sydney 2052, Australia}
\author{V. V. Flambaum}
\affiliation{School of Physics, University of New South Wales,
Sydney 2052, Australia}

\date{\today}

\begin{abstract}

We present an efficient method  of inclusion of the core-valence 
correlations into the configuration interaction (CI) calculations.
These correlations take place in the core area where the
potential of external electrons is
 approximately constant. A constant potential does not change
the core electron wave functions and Green's functions.
Therefore, all operators describing interaction of $M$ valence electrons
and $N-M$ core electrons (the core part of the Hartree-Fock Hamiltonian
 $V^{N-M}$,
the correlation potential $\hat\Sigma_1({\bf r},{\bf r'},E)$ and the
screening of interaction between valence electrons by the core electrons
$\hat\Sigma_2$) may be calculated with all
 $M$ valence electrons removed.
This allows one to avoid subtraction diagrams which make accurate inclusion
of the core-valence correlations for $M>2$ prohibitively complicated.
Then the CI Hamiltonian for $M$ valence
electrons  is calculated using
orbitals in complete $V^{N}$ potential (the mean field produced by all
 electrons); $\hat\Sigma_1$ + $\hat\Sigma_2$ are added to the CI Hamiltonian
to account for the core-valence correlations.
We calculate $\hat\Sigma_1$ and  $\hat\Sigma_2$ 
using many-body perturbation theory in which dominating classes
of diagrams are included in all orders.
 We use neutral  Xe~I and all positive ions up to
Xe~VIII as a testing ground. We found that the core electron density for
all these systems is practically the same. Therefore, we use
the same  $\hat\Sigma_1$ and  $\hat\Sigma_2$ to build the CI
Hamiltonian in all these systems ($M=1,2,3,4,5,6,7,8$).
Good agreement with experiment 
for energy levels and Land\'{e}
factors is demonstrated for all cases from Xe~I to
Xe~VIII.

\end{abstract}


\maketitle

\section{Introduction}

Accurate calculations for many-electron atoms play an important role 
in many advanced topics of modern physics. This includes parity and
time invariance violating phenomena in atoms~\cite{Ginges}, search 
for manifestation of possible variation of fundamental constants in 
astrophysical data~\cite{alpha}, or in present-day laboratory 
experiments~\cite{clock}, improving accuracy of atomic clocks~\cite{BB},
study of super-heavy elements (see, e.g.~\cite{super}), etc. 
Calculations are needed for planning of experiments and interpretation
of the results.

Atoms of the most interest for the listed topics are usually found in 
the second part of the periodic table where measurements or observations
are more likely to give useful information due to strong enhancement of
the effects caused by interplay between relativistic and many-body effects.
On the other hand, accurate treatment of relativistic and many-body
effects represent a big challenge for atomic calculations. Not surprisingly,
the number of methods capable of producing reliable and accurate results
is very limited. The most advanced methods have been developed for atoms
with one external electron above closed shells. For example, most accurate
calculations of the parity non-conservation in cesium where carried out 
with two most advanced methods. One was
the correlation potential (CP) method~\cite{CPM} combined with the 
all-order perturbation theory in screened Coulomb interaction~\cite{DSF89}
and the other was
the linearized coupled cluster approach~\cite{Blundell90}.
With these two methods, energy levels and transition amplitudes for
alkali atoms can be calculated to the accuracy of fraction of percent
while hyperfine structure and parity violating amplitudes are calculated
to the accuracy of 0.5 - 1\%~\cite{DSF89,Blundell90,CsPNC}.

For atoms with more than one external electron in open shells the 
accuracy of calculations is significantly lower. For example, the
best accuracy achieved for PNC in Tl is around 3\% (3\% in Ref.~\cite{TlPNC}
and 2.5\% in Ref.~\cite{Kozlov01}). Typical accuracy for energies is
about 1\% or worse. The main challenge is the need for accurate treatment
of both core-valence and valence-valence correlations.
The most commonly used methods can be divided in several main groups:
(a) many-body perturbation theory (MBPT) (see, e.g.~\cite{MBPT}),
(b) coupled cluster approach (CC) (see, e.g.~\cite{CC2}),
(c) configuration interaction (CI) (see, e.g.~\cite{CI}),
and (d) multi-configuration Dirac-Fock method (MCDF)(see, e.q.~\cite{MCDF}).
There are also combinations of these basic techniques.

All of these method have their limitations. For example, CI usually
treats correlations between valence electrons very accurately but
core-valence correlations are either totally neglected or only
small fraction of them is included. MBPT can include more core-valence
correlations, but its application to the correlations between valence
electrons is limited by the fact that these correlations are often
too strong to be treated perturbatively. The CC approach includes
certain types of core-valence and valence-valence correlations in
all-orders and in principle can be formulated for any number of
valence electrons. However, the equations are complicated and
most of practical realization of the method deal with only one 
or two electrons (or an electron and a hole).

Significant progress can be achieved by combining different techniques.
In 1996 we developed a method which combines the MBPT with the CI 
method (CI+MBPT)~\cite{CI+MBPT}. Here, the second-order MBPT was used to
construct the effective Hamiltonian in the valence space which includes 
the core-valence correlations. It differs from the standard CI Hamiltonian
by an extra correlation operator $\hat \Sigma$ which accounts for
the core-valence correlations. Single-electron part of this operator
is very similar to the correlation potential used for atoms with
one external electron~\cite{CPM}.
It was demonstrated that inclusion of the core-valence correlations lead
to a significant improvement of the accuracy of calculations 
(see, e.g.~\cite{CI+MBPT,Ba,hfs,Savukov02}). Savukov and 
co-workers~\cite{Savukov02a} developed 
a version of the method which uses the hole-particle formalism. They
applied the technique for a calculation of the electron structure
of the noble-gas atoms~\cite{Savukov02a,Savukov03}.

The CI+MBPT method was successfully used for a number of atoms with 
two or three valence electrons~\cite{CI+MBPT,Ba,Ra} (or an electron and a 
hole~\cite{Savukov02a,Savukov03}). 
Its extension to atoms with more electrons in open
shells meets some difficulties. It turns out that convergence of the
MBPT varies very much from atom to atom and strongly depends on an
initial approximation. The core-valence correlations are often 
too large if treated in the same fashion as in our original works 
and their inclusion does not improve the results. 

It is widely accepted that the Hartree-Fock potential is the best
choice as a zero approximation for consequent use of the MBPT due to great 
reduction of the terms caused by exact cancellation between the potential 
and electron-electron Coulomb terms.
However, for atoms with open shells HF procedure is not defined
unambiguously. This is especially true when the CI method is to be used.
Here we have freedom of how many electrons are to be included into the initial
HF procedure and how many electrons are to be treated as valence electrons
in the CI calculations. It was found in our previous work ~\cite{vn}
that for a wide range of atoms the best choice is the so-called
$V^{N-M}$ approximation. These are the atoms in which valence electrons
form a separate shell, defined by the same principal quantum number.
 For example,
the ground state configuration of xenon is [Pd]$5s^25p^6$. Its eight outermost
electrons have $n=5$ while all other electrons have $n<5$. This means that
eight outermost electrons should be treated as valence electrons and the
initial HF procedure should not include them. This greatly simplifies
the MBPT, improves its convergence and allows one to include higher-order
correlations in the same way as it was done for atoms with one external
electron. 

 The aim of the present work is to develop a solid theoretical background
for use of the $V^{N-M}$ potential as a starting point. In principle,
 this starting point is  equivalent to any other choice of the
initial HF potential. Indeed, the actual role of the 
{\it subtraction diagrams} in the correlation operators $\hat \Sigma$
is to reduce results obtained with any zero approximation
 to $V^{N-M}$ results (see an explanation below).
 However, the technique for   $V^{N-M}$ is much simpler
(no subtraction diagrams) and allows us to sum dominating chains of higher
order diagrams to all orders (it is practically impossible for other choice
 of zero approximation). This results in a higher accuracy.
 Another advance of the present work is the use of a compact basis for valence
states. In our previous works~\cite{Ra,vn,vn1,vn2} we used the same basis
 to calculate 
$\hat \Sigma$ and to do the CI calculations. The basis was formed from
the eigenstates of the $\hat V^{N-M}$ potential. This had an 
advantage of having the same single- and double-electron matrix elements
for all ions of the same atom. Moving from ion to ion was easy, requiring
only change the number of electrons in the CI calculations. However,
convergence of the CI calculations rapidly deteriorated with growing number
of electrons. When number of electrons became as large as eight, saturation
of the basis was very hard to achieve unless huge computer resources were used.
In present work we demonstrate that the basis states for valence electrons
don't have to be eigenstates of the $\hat V^{N-M}$ potential. Instead,
HF states calculated in the mean field of all electrons
(of a neutral atom or corresponding positive ion) can be used
after minor modifications. In this case we have
to recalculate the CI basis when we change the number
of valence electrons $M$. However, the gain is much larger. Since HF states are
already good approximations to the wave functions of valence electrons
we can limit the basis to just few states in each partial wave. Therefore,
even for eight valence electrons the CI matrix is small, its calculation 
and diagonalization takes little time but the final results are very accurate.

We calculate energy levels and Land\'{e} $g$-factors for neutral xenon
and all its positive ions from X~II to Xe~VIII for illustration on how
the technique works. Good agreement with experiment is demonstrated 
for all cases while very little computer resources are needed on every
stage of the calculations.

\section{Core electron density and potential in  $V^{N-M}$ and $V^{N}$
 approximations}

The effective Hamiltonian of the CI method has the form 
(see, e.g.~\cite{CI+MBPT})
\begin{equation}
  \hat H^{\rm eff} = \sum_{i=1} \hat h_{1i} + \sum_{i < j} 
\frac{e^2}{|\mathbf{r}_i - \mathbf{r}_j|}.
\label{heff}
\end{equation}
Summation goes over valence electrons,
$\hat h_1(r_i)$ is the one-electron part of the Hamiltonian
\begin{equation}
  \hat h_1 = c \mathbf{\alpha p} + (\beta -1)mc^2 - \frac{Ze^2}{r} + V_{core},
\label{h1}
\end{equation}
$\mathbf{\alpha}$ and $\beta$ are the Dirac matrices, $Ze$ is the 
nuclear charge and
$V_{core}$ is the electrostatic potential created by the core electrons.
Regardless of initial approximation used to calculate core and valence
states, the valence electrons never contribute to $V_{core}$ directly.
They can only contribute to $V_{core}$ via the self-consistent HF procedure or
via any other potential used to represent valence electrons.
If the core electrons and valence electrons belong to different shells
  the effect of the valence electrons  on electron
states in the core and thus on $V_{core}$ can be extremely small.
Indeed, in this case the overlap between density of the valence electrons
and density of the core electrons is small. Therefore, the exchange
interaction between the core and valence electrons, which is proportional
to the overlap, is negligible in comparison with energy of the core electrons.
 On the other hand, the direct potential created 
by the valence electrons is practically constant inside the core since
nearly all charge of the valence electrons is located outside the core.
 Constant potential corresponds
to zero electric field and cannot have any effect on the wave function
of the core electrons. The only effect of the constant potential $V_0$
is in energy shift $\delta E=V_0$. However, it does not change the
single-particle wave functions and Green's functions of core electrons
 since the wave equation contains the difference $E-V_0$ which does not change.
We may formulate this conclusion using the perturbation theory.
In first order in $V_0$ a  core state $a$  in the $V^{N-M}$ approximation
 and $\tilde a$ in the $V^N$ approximation are related by
\begin{equation}
  |\tilde a \rangle = |a \rangle + \sum_n \frac{\langle a|V_0|n \rangle}
{E_a - E_n} | n \rangle.
\label{avn}
\end{equation}
If the potential $V_0$ is constant, the matrix element 
 $\langle a|V_0|n \rangle=V_0\langle a|n \rangle=0$ due to the orthogonality
condition. This explains why the changes of the core wave functions,
  density and  potential are very small. 

\begin{figure}
\centering
\epsfig{figure=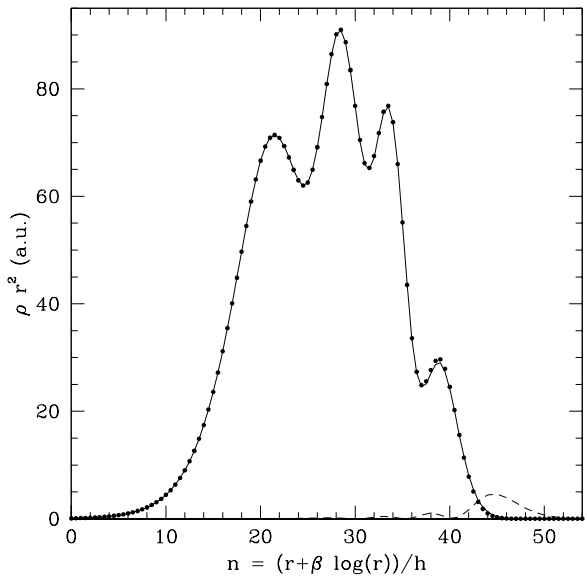,scale=0.4}
\caption{Electron densities (multiplied by $r^2$) of Xe~I and Xe~IX. 
Atomic core ($n$=1,2,3,4) of Xe~I, solid line; $5s$ and $5p$, dashed line; 
electron density of Xe~IX, dots.}
\label{xe-fig}
\end{figure}

Small overlap between the core and valence states usually takes place when
 these
states correspond to different atomic shells defined by the principal
quantum number (see, e.g.~\cite{vn,vn1,vn2}). In case of xenon, eight 
outermost electrons have principal quantum number $n=5$ (the $5s^25p^6$ 
ground state configuration), while all core electrons have $n<5$.
Therefore, if the eight electrons are considered to be valence electrons
we should expect that they have little effect on the core states.
Fig.~\ref{xe-fig} shows electron densities of Xe~I and Xe~IX calculated
in the HF approximation. For the neutral Xe~I electron densities of valence
and core electrons are shown separately. One can see that the overlap between
them is indeed very small. Therefore, it turns out that when electron density
of Xe~IX is calculated it practically coincides with the electron density
of the core states of neutral Xe. The former is shown by dots of
Fig.~\ref{xe-fig}. Resolution of this figure doesn't allow us to see any 
difference between electron densities of Xe~IX and the core of Xe~I.
This is in spite of huge difference in energies of core states of two
atoms. One may argue that huge difference in energies should lead to
a noticeable difference in wave functions, at least on large distances.
Indeed, a wave function of an atomic electron has asymptotic defined by its
energy and potential
\begin{equation}
  \psi(r) \sim e^{-\int \sqrt{2m (E-V(r))}dr}.
\label{assimp}
\end{equation}
However, in the area up to the radius of the valence shell
 there is actually no difference in $E-V(r)$ for the core orbitals
in Xe~IX and Xe~I since $\delta E=\langle \delta V \rangle $. 
The difference in asymptotic behavior appears only
near the radius of the valence shell where the core electron density
is extremely small.

Thus we conclude that the core electron density and potential
have practically no dependence on the number of valence electrons
if the valence electrons are in a different shell.


\section{The core-valence correlation corrections in
 the $V^{N}$ and $V^{N-M}$ approximations}

The use of the Feynman diagram technique allows us to express
the core-valence correlation corrections in terms of the single-particle
wave functions and Green's functions \cite{DSF89} (see Appendix).
Therefore, all the arguments presented above are  applicable when we
consider calculation of the correlation operators
$\hat \Sigma_1$ and  $\hat \Sigma_2$; they may be calculated using $V^{N-M}$
basis for core electrons.
 
   It may be instructive to clarify this conclusion using  more popular
 Schroedinger perturbation theory where explicit summation over intermediate
states is involved. The correlations between the valence and core electrons
as well as the screening of the interaction between the valence electrons
happen inside the area occupied by the core electrons. Let us enclose
the core  by a sphere with zero boundary condition for the core electrons.
This allows us to reduce the core electron problem to the discrete spectrum.
Let us now consider the interaction of the core electrons with external
 electrons using the perturbation theory. The constant potential $V_0$ of
 external electrons does not change the core electron wave functions.
 It also does not change the energy differences $E_n-E_m$ between the enclosed
 ``core''states, they are shifted by the same energy $V_0$ (note that these
enclosed  states form complete basis set inside the sphere). Therefore, all
 the terms in the perturbation
theory for the core-valence interaction (beyond the mean field which we take
into account in the $V^N$ valence orbitals) do not depend on the spectator
 valence electrons. This is why we can calculate all core-valence
correlations using the $V^{N-M}$ core orbitals. To avoid misunderstanding
 we should note that we use this picture for the explanation only,
 no special boundary
conditions  for core electrons are needed for actual calculations
(it is obvious if we use the Green's function technique;
all the integrals over coordinates are dominated by the core
area where the correlations between the valence and core electrons
actually happen).  Note  that below we do not neglect effects of $V_0$,
 we only treat them as a perturbation since the non-diagonal matrix elements   
 $\langle a|V_0|n \rangle$ are small.

The effective Hamiltonian of the CI+MBPT method has the form similar to
(\ref{heff}) but with extra terms for single and double electron parts of it.
These terms, for which we use notation $\hat \Sigma$, describe the core-valence
correlations \cite{CI+MBPT}. There is a single-electron operator 
$\hat \Sigma_1$ which is added to the single-electron part $\hat h_1$ 
(\ref{h1}) of the CI Hamiltonian:
\begin{equation}
  \hat h_1(r) \rightarrow \hat h_1 + \hat \Sigma_1 .
\label{h1n}
\end{equation}
$\hat \Sigma_1$ describes correlations between a particular valence electron
and core electrons. It is very similar to the correlation potential 
$\hat \Sigma$ used for atoms with one external electron (see, e.g.~\cite{CPM}).

There is also a two-electron operator $\hat \Sigma_2$ which modifies
Coulomb interaction between valence electrons:
\begin{equation}
  \frac{e^2}{|\mathbf{r_1 - r_2}|} \rightarrow \frac{e^2}{|\mathbf{r_1 -
      r_2}|} 
+ \hat \Sigma_2.
\label{h2n}
\end{equation}
Physical interpretation of $\hat \Sigma_2$ is the screening of Coulomb
 interaction
between valence electrons by core electrons.

When number of valence electrons is greater than 2 there is also a three-body
operator $\hat \Sigma_3$  \cite{CI+MBPT} and higher-order many-body
 operators $\hat \Sigma_4$, $\hat \Sigma_5$, etc.    .
 However, they are usually very
small and we will not consider them in the present work.

\begin{figure}
\centering
\epsfig{figure=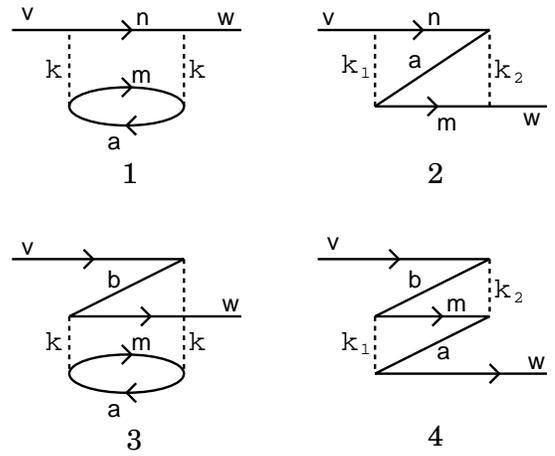,scale=0.8}
\caption{Second-order diagrams for single-electron correlation
operator $\hat \Sigma^{(2)}_1$.}
\label{sigma1-fig}
\end{figure}

\begin{figure}
\centering
\epsfig{figure=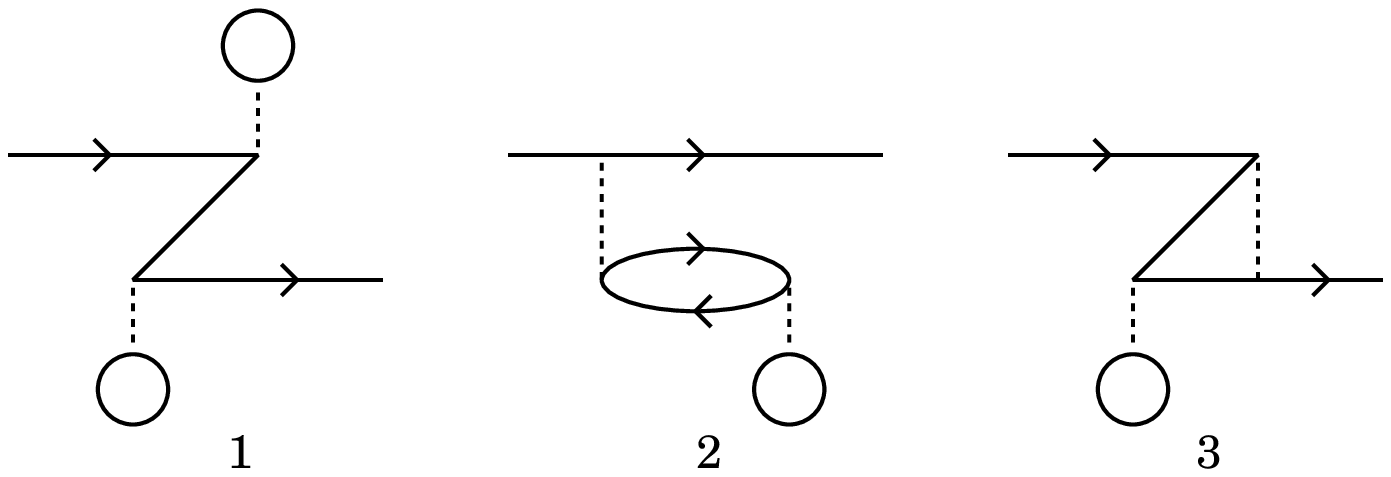,scale=0.6}
\caption{Subtraction diagrams for $\hat \Sigma^{(2)}_1$.}
\label{sigma1sub-fig}
\end{figure}

\begin{figure}
\centering
\epsfig{figure=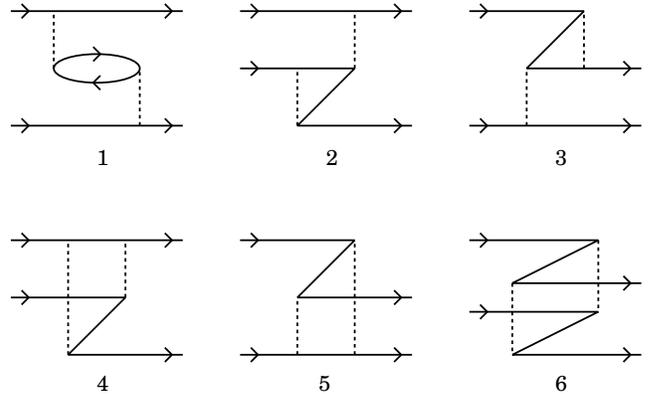,scale=0.6}
\caption{Second-order diagrams for double-electron correlation
operator $\hat \Sigma^{(2)}_2$.}
\label{sigma2-fig}
\end{figure}

\begin{figure}
\centering
\epsfig{figure=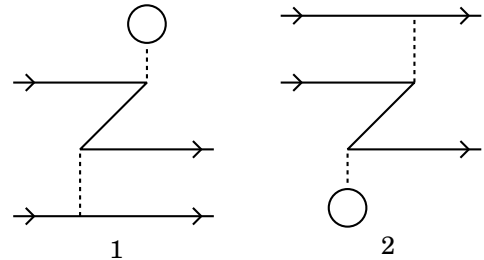,scale=0.7}
\caption{Subtraction diagrams for $\hat \Sigma^{(2)}_2$.}
\label{sigma2sub-fig}
\end{figure}

The  full set of diagrams for $\hat
\Sigma_1$ and $\hat \Sigma_2$ in the second order of MBPT is presented on
Figs. \ref{sigma1-fig}, \ref{sigma1sub-fig}, \ref{sigma2-fig} and 
\ref{sigma2sub-fig}. It contains the so-called {\it subtraction} diagrams
which are proportional to $V_{core} - V^{HF}$, where $V_{core}$ is the 
potential of the core electrons as in the effective CI Hamiltonian
(\ref{heff}), $V^{HF}$ is the potential in which states of the core were
calculated. Note that subtraction diagrams vanish in the $V^{N-M}$
approximation: $V_{core} = V^{HF}$.

 The origin of the subtraction
diagrams is clear from the definition of the perturbation
(residual interaction) operator
$U=H_{exact}-H_0$ where $H_{exact}$ is the exact Hamiltonian
 and $H_0$ is the zero
 approximation Hamiltonian. If the field of external electrons
is included in $H_0$ it produces additional
contributions which we call the subtraction diagrams.
Thus, the potential $V_0$ appears with positive sign in the mean
field (and core wave functions), and with negative sign in the
 residual interaction (and subtraction diagrams). If we calculate
all correlations exactly, to all orders, $V_0$ must disappear in the final
result. In any finite order of the many-body perturbation theory
there are only partial cancellations;
 lower orders of expansion in $V_0$ are canceled out.  
Thus, the role of the subtraction diagrams 
is to cancel  the potential of spectator valence electrons
acting on the core electrons (effect of  valence electrons
on the  core lines of the diagrams).
In other words, 
the subtraction diagrams guarantee that in any given order of
expansion in $V_0$ 
the operators  $\hat\Sigma_1$ and $\hat \Sigma_2$
are reduced to the results of $V^{N-M}$ approximation.
Therefore, if the non-diagonal matrix elements of $V_0$ are small
the  $V^{N-M}$ approximation is the best zero approximation
since the calculations are much simpler (no subtraction diagrams).

  It is easy  to see all these cancellations of $V_0$
explicitly, order by order in $V_0$. Here one should remember
that  change of the  valence electron energies due to
change of the core Hartree-Fock potential
 (which formally has the first order in the Coulomb interaction)
 is actually canceled
by the second order subtraction diagrams; contribution of $V_0$
into the core wave functions in the second order diagrams
is canceled by the third order subtraction diagrams, etc.  
 
\section{Higher order terms in $\hat \Sigma$}

We have seen above that if the electrostatic potential $V_0$ created by valence
electrons is nearly constant inside the core then the $V^{N-M}$ approximation
is equivalent to the $V^{N-M}+V_0$ approximation, where $V_0$ can have
contributions from all $M$ valence electrons, or any fraction on them or it
can be just a model potential. The only condition is that $V_0$ is nearly
constant inside the core. This means that without any compromise on accuracy
we can do the calculation in the $V^{N-M}$ approximation which is technically
more simple. Another advantage of using $V^{N-M}$ is that we have the same
effective Hamiltonian for any number of valence electrons from 1 to
$M$. Therefore we can do the calculations in a very similar way for all
corresponding ions as well as for a neutral atom.

Eliminating subtracting diagrams in the $V^{N-M}$ approximation makes $\hat
\Sigma_1$ practically identical to the correlation potential $\hat \Sigma$ used
for atoms with one external electron. Therefore, we can try to improve the
accuracy of calculations by including important higher order terms into $\hat
\Sigma_1$ the same way as it was done in a number of calculations for alkali
atoms (see, e.g.~\cite{DSF89,CsPNC}). We include two dominating classes of 
higher order diagrams into calculation of $\hat \Sigma_1$. One is screening 
of the Coulomb interaction between valence and core electrons by other core
electrons. Another is interaction between an electron excited from the core
and a hole in the core created by this excitation. Both classes of diagrams
are included in all orders (see, e.g.~\cite{DSF89,CsPNC} for details).

\begin{table}
  \caption{Removal energies of lowest states of Xe~VIII (cm$^{-1}$) in
different approximations; comparison with experiment}
  \label{XeVIII}
  \begin{ruledtabular}
    \begin{tabular}{c c c c c}
      State & HF &  $\hat \Sigma^{(2)}$ 
            &  $\hat \Sigma^{(\infty)}$ & Expt~\cite{NIST} \\
            \hline
  $5s_{1/2}$ &   839764 &  858722 &   854842 &   854755 \\
  $5p_{1/2}$ &   725342 &  741343 &   738280 &   738288 \\
  $5p_{3/2}$ &   707377 &  722378 &   719550 &   719703 \\
  $5d_{3/2}$ &   536494 &  546174 &   544092 &   544867 \\
  $5d_{5/2}$ &   533632 &  543150 &   541096 &   541939 \\
  $4f_{5/2}$ &   572050 &  590725 &   587324 &   589594 \\
  $4f_{7/2}$ &   571684 &  590056 &   586717 &   589044 \\
    \end{tabular}
  \end{ruledtabular}
\end{table}

We use notation $\hat \Sigma^{(\infty)}_1$ for the all-order $\hat \Sigma$
operator as compared to $\hat \Sigma^{(2)}_1$ for the second-order
operator. The effect of inclusion of second and higher order correlations can
be illustrated by calculating of the energy levels of Xe~VIII. This ion has
only one valence electron and calculations for it can be done the same way as
for other single-valence electron atoms (see, e.g.~\cite{DSF89}). Instead of
diagonalizing the CI matrix we solve HF-like equations for valence electrons
in coordinate space, with $\hat \Sigma_1$ included in it:
\begin{equation}
  (h_1 + \hat \Sigma_1 - \epsilon) \psi = 0.
  \label{Brueck}
\end{equation}
Here single-electron Hamiltonian $h_1$ is given by (\ref{h1}) while $\hat
\Sigma_1$ can be either $\hat \Sigma^{(2)}_1$ or $\hat\Sigma^{(\infty)}_1$. If
no $\hat \Sigma_1$ is included then eq. (\ref{Brueck}) gives HF energies and
wave functions. 

The results of calculations are presented in Table~\ref{XeVIII} and compared
with experiment. One can see systematic significant improvement of the
results when first $\hat \Sigma^{(2)}_1$ and then $\hat \Sigma^{(\infty)}_1$
are included.

We are now going to use the same $\hat \Sigma^{(\infty)}_1$ operator for all
ions from Xe~VII to Xe~II and for neutral xenon. For all these ions which have
more than one valence electrons the $\hat \Sigma_2$ operator should also be
included. In the $V^{N-M}$ approximation the $\hat \Sigma_2$ term is given by
diagrams on Fig.~\ref{sigma2-fig}, and no subtraction diagrams are needed.
To include higher-order correlations into $\hat \Sigma_2$ we use screening
factors the same way as we do this for the exchange diagrams of $\hat
\Sigma_1$ (Figs.~\ref{sigma1-fig}.2 and \ref{sigma1-fig}.4) 
(see, e.g.~\cite{DSF89}). To explain how
screening factors are found and used we need to go into more details on how
the all-order correlation operator $\hat \Sigma^{(\infty)}_1$ is
calculated. We use Feynman diagram technique to calculate {\em direct} diagrams
(Fig.~\ref{sigma1-fig} 1 and 3). It allows us to include an infinite chain of
screening diagrams in all orders~\cite{DSF89}. Application of the Feynman diagram
technique to {\em exchange} diagrams (Fig.~\ref{sigma1-fig} 2 and 4) is much
more complicated~\cite{CsPNC}. On the other hand these diagrams are usually an
order of magnitude smaller than {\em direct} diagrams. Therefore it makes
sense to use an approximate method by introducing {\em screening factors}.
We assume that screening of Coulomb interaction between core and valence
electrons depends only on multipolarity $k$ of Coulomb interaction. Screening
factors $f_k$ are calculated as ratios of partial contributions to $\hat
\Sigma_1$:
\begin{equation}
  f_k = \langle \hat \Sigma^{(\infty)}_k \rangle / \langle \hat
  \Sigma^{(2)}_k \rangle,
  \label{fk}
\end{equation}
where only {\em direct} diagrams are included in $\hat \Sigma^{(2)}_k$
and $\hat \Sigma^{(\infty)}_k$ and only screening of Coulomb interaction but
no hole particle interaction is included in $\hat \Sigma^{(\infty)}_k$. The
values of $f_k$ found from calculations for alkali atoms are

\begin{equation}
\begin{array}{llll}
  f_0 = 0.72, & f_1 = 0.62, & f_2 = 0.83, & \\ 
  f_3 = 0.89, & f_4 = 0.94, & f_5 = 1.00, &  \rm{etc.} \\
\end{array}
  \label{fkv}
\end{equation}
The values of $f_k$ change very little from atom to atom and the values 
presented above can be used for xenon. This is supported by the results
obtained for Xe~VIII (see Table~\ref{XeVIII}).

The effect of $\hat \Sigma_2$ on atomic energies is much smaller than those
of $\hat \Sigma_1$. Therefore we can also treat higher-order correlations in
$\hat \Sigma_2$ in an approximate way, via screening factors, as we do this 
for exchange part of $\hat \Sigma_1$. We replace every Coulomb integral $Q_k$
on all diagrams on Fig.~\ref{sigma2-fig} except diagram
Fig.~\ref{sigma2-fig}.1 by its screened values $f_k Q_k$ where screened
factors $f_k$ are taken as in (\ref{fkv}). For the diagram
Fig.~\ref{sigma2-fig}.1 only one of the Coulomb integrals is replaced by its
screened value. This is because this diagram can generate only one infinite
chain of loops representing screening. Therefore, screening should by included
only once. This this very similar to the all-order treatment of the direct
diagram for $\hat \Sigma_1$ (Fig.~\ref{sigma1-fig}.1 and \ref{sigma1-fig}.3). 
If this diagram is expressed in terms of screened Coulomb interaction, 
only one of two Coulomb integrals should be replaced by a screened one 
(see~\cite{DSF89,CsPNC} for details).

\section{Basis}

\begin{table}
  \caption{Basis states of valence electrons used in the CI calculations,
their total number ($N$) for each atom or ion
and HF configurations in which they were calculated}
  \label{basis}
  \begin{ruledtabular}
    \begin{tabular}{l r l l l}
      \multicolumn{1}{c}{Atom} & \multicolumn{1}{c}{$N$} &
      \multicolumn{1}{c}{Basis states} &
      \multicolumn{2}{c}{Configurations} \\ 
\hline
Xe~I   & 15 & 5s,6s,7s,5p,6p,7p,5d,6d,4f  & $5s^25p^6$, & $5s^25p^5nl$ \\
Xe~II  & 10 & 5s,6s,5p,6p,5d,4f           & $5s^25p^5$, & $5s^25p^4nl$ \\
Xe~III &  8 & 5s,6s,5p,6p,5d              & $5s^25p^4$, & $5s^25p^3nl$ \\
Xe~IV  & 10 & 5s,6s,5p,6p,5d,4f           & $5s^25p^3$, & $5s^25p^2nl$ \\
Xe~V   &  8 & 5s,6s,5p,6p,5d              & $5s^25p^2$, & $5s^25pnl$ \\
Xe~VI  &  8 & 5s,6s,5p,6p,5d              & $5s^25p  $, & $5s^2  nl$ \\
Xe~VII &  8 & 5s,6s,5p,5d                 & $5s^2    $, & $5s    nl$ \\
    \end{tabular}
  \end{ruledtabular}
\end{table}

There are two single-electron basis sets in this problem. One is used to 
calculate $\hat \Sigma$ and other is used to construct many-electron states
of valence electrons for the CI calculations.

 In principle, it is possible to use the same
basis for both purposes and we did so in many of our earlier
calculations~\cite{Ba,Ra,vn,vn1,vn2}. The most convenient choice for the basis is the basis
consisting of single-electron states calculated in the $V^{N-M}$ potential. 
We use B-spline technique to calculate the basis. Lower and upper component of
each basis set is expressed as linear combination of 40 B-splines in the
cavity of radius of $40a_B$. Expansion coefficients are found from the condition
that the basis states are the eigenstates of HF Hamiltonian (\ref{h1}) with
the $V^{N-M}$ potential. The advantages of this approach are many:
core and valence states are orthogonal automatically, the basis is reasonably
complete and does not depend on number of valence electrons. The latter means
in particular that if we want to change number of valence electrons (e.g. to
do calculation for another ion) we don't have to recalculate single and double
electron matrix elements. The shortcoming of this approach is rapid increase
of the size of the CI matrix with the number of valence electrons. Indeed,
typical number of single-electrons basis states needed to get saturation of
the basis is around 100. The number of ways, valence electrons can be
distributed over these 100 states grows very fast with the number of valence
electrons. For eight electrons like for Xe~I the matrix reaches unmanageable
size, even when some configuration selection technique is used. 

In present work we use the basis described above only for calculation of $\hat
\Sigma$. For the CI calculations we use very compact basis of HF states of
corresponding ion or neutral atom. For example, we perform HF calculations for
neutral Xe~I in its ground state [Pb]$5s^25p^6$ in the $V^N$ approximation and
then use the $5s$ and $5p$ states as the basis states for the CI calculations
for Xe~I in the $V^{N-8}$ approximation. Other basis states like $6s$, $6p$,
etc. are obtained by removing one $5p$ electron from the atom and calculating
these states in the frozen field of remaining electrons. The states
obtained this way are not orthogonal to the core which corresponds to the
$V^{N-8}$ potential. In the relativistic case the basis
 states for valence
electrons must also be orthogonal  to the negative
energy states (positron states). Both conditions (orthogonality to the core
and to the negative energy states) can be achieved by projecting
a basis state on the B-spline states above the core: 

\begin{equation}
  | v \rangle \rightarrow | v'  \rangle = \sum_i | i \rangle \langle i | v \rangle. 
  \label{proj}
\end{equation}
Here summation goes over states above the core. Functions $| v' \rangle$ are
more suitable for the CI calculations than states $| v \rangle$ because they
don't have admixture of the core and negative energy states.

If more than one state of particular symmetry is included into the basis (like,
e.g. the $6p$ and $7p$ states for Xe~I) they also need to be orthogonalized to
each other.

Full list of valence states for xenon and its ions used in the calculations
are presented in Table~\ref{basis}. First column shows an atom or ion, second
column gives total number of valence basis states, then states are listed
together with the configurations in which they were calculated.
 Note that every state with $l>0$
consists of two function, e.g. $6p$ stands for $6p_{1/2}$ and $6p_{3/2}$, etc.
Note also that the number of basis states is always small, much smaller than
about 100 needed with the B-spline basis. This greatly overweights an
inconvenience of recalculating the basis for every ion or atom.

\section{Calculations for xenon and its ions}

In this section we present calculations for xenon and its ions. The whole
calculation scheme consists of the following steps (we use Xe~I as an example):
\begin{enumerate}
\item HF for Xe~IX, $V^{N-8}$ potential is obtained.
\item Calculation of B-spline states in the $V^{N-8}$ potential.
\item Calculation of $\hat \Sigma_1$.
\item HF for Xe~I, the $5s$ and $5p$ basis states are obtained.
\item Calculation of valence basis states.
\item Calculation of single and double-electron matrix elements, including
  matrix elements of $\hat \Sigma_2$.
\item Calculation and diagonalization of the CI matrix.
\end{enumerate}
Ar first glance this scheme doesn't look very simple. However, none of the
steps listed above are very time consuming or require large computer
power. The most time consuming step is calculation of $\hat \Sigma$ ($\hat
\Sigma_1$ in step 3 and $\hat \Sigma_2$ in step 6). An efficient way of
calculating both $\hat \Sigma_1$ and $\hat \Sigma_2$ is presented in the
appendix. The timescale to obtain all results presented in this section while
using a PC or a laptop is one day.
 
\begin{table*}
  \caption{Ground state removal energy (a.u.), excitation energies
(cm$^{-1}$) and $g$-factors of lowest states of Xe in different
approximations}
  \label{Xe}
  \begin{ruledtabular}
    \begin{tabular}{l l c c c c c c l c c l r}
      \multicolumn{2}{c}{State} & $J$ & \multicolumn{1}{c}{CI} &
      \multicolumn{1}{c}{$\hat \Sigma_1^{(2)}$} &
      \multicolumn{1}{c}{$\hat \Sigma_1^{(2)} \& \hat \Sigma_2^{(2)}$} &
      \multicolumn{1}{c}{$\hat \Sigma_1^{(\infty)}$} &
      \multicolumn{1}{c}{$\hat \Sigma_1^{(\infty)} \& \hat \Sigma_2^{(2)}$} &
      \multicolumn{2}{c}{$\hat \Sigma^{(\infty)}$} &
      \multicolumn{2}{c}{Expt} & \\
 & & & \multicolumn{1}{c}{$E$} & \multicolumn{1}{c}{$E$} & 
       \multicolumn{1}{c}{$E$} & \multicolumn{1}{c}{$E$} & \multicolumn{1}{c}{$E$} & 
       \multicolumn{1}{c}{$E$} & \multicolumn{1}{c}{$g$} &
       \multicolumn{1}{c}{$E$} & \multicolumn{1}{c}{$g$} & 
       \multicolumn{1}{c}{$\Delta$} \\
\hline
$5s^25p^6$   & $^1$S         & 0 & -15.21 & -15.76 & -15.69 &-15.53 & -15.48 & -15.49 &         & -15.61&        &      \\
\hline
$5s^25p^56s$ & $^2$[3/2]$^o$ & 2 &  62710 &  70595 &  68587 & 68289 &  66310 &  67040 &  1.4994 &  67068&  1.50095&    28 \\
             &               & 1 &  64013 &  71916 &  69873 & 69594 &  67573 &  68319 &  1.2157 &  68045&  1.2055 &  -274 \\
$5s^25p^56s$ & $^2$[1/2]$^o$ & 0 &  71616 &  80192 &  78425 & 77568 &  75824 &  76480 &  0      &  76197&         &  -283 \\
             &               & 1 &  72896 &  81586 &  79764 & 78925 &  77120 &  77799 &  1.3160 &  77185&  1.321  &  -614 \\
$5s^25p^56p$ & $^2$[1/2]     & 1 &  72707 &  81332 &  79318 & 78519 &  76533 &  77283 &  1.8559 &  77269&  1.852  &   -14 \\
             &               & 0 &  76219 &  84202 &  82183 & 81602 &  79607 &  80350 &  0.0000 &  80119&         &  -231 \\
$5s^25p^56p$ & $^2$[5/2]     & 2 &  73810 &  82369 &  80304 & 79587 &  77545 &  78307 &  1.1005 &  78120&  1.11103&  -187 \\
             &               & 3 &  74045 &  82627 &  80568 & 79837 &  77802 &  78564 &  1.3333 &  78403&  1.336  &  -161 \\
$5s^25p^56p$ & $^2$[3/2]     & 1 &  74755 &  83339 &  81261 & 80549 &  78496 &  79260 &  1.0232 &  78956&  1.02348&  -304 \\
             &               & 2 &  74964 &  83540 &  81466 & 80755 &  78705 &  79468 &  1.3913 &  79212&  1.3836 &  -256 \\
$5s^25p^55d$ & $^2$[1/2]$^o$ & 0 &  76068 &  84821 &  82949 & 82082 &  80235 &  80919 &  0      &  79771&         & -1148 \\
             &               & 1 &  76259 &  84946 &  83067 & 82221 &  80367 &  81054 &  1.3786 &  79987&  1.395  & -1067 \\
$5s^25p^55d$ & $^2$[7/2]$^o$ & 4 &  76425 &  84622 &  82869 & 81997 &  80257 &  80900 &  1.2500 &  80197&  1.2506 &  -703 \\
             &               & 3 &  77283 &  85513 &  83764 & 82877 &  81141 &  81783 &  1.0762 &  80970&  1.0749 &  -813 \\
$5s^25p^55d$ & $^2$[3/2]$^o$ & 2 &  76370 &  84667 &  82973 & 82001 &  80323 &  80947 &  1.3775 &  80323&  1.3750 &  -624 \\
             &               & 1 &  80595 &  89175 &  86989 & 86445 &  84279 &  85087 &  0.9900 &  83890&         & -1197 \\
$5s^25p^55d$ & $^2$[5/2]$^o$ & 2 &  78310 &  86627 &  84853 & 83959 &  82199 &  82846 &  0.9419 &  81926&         &  -920 \\
             &               & 3 &  78626 &  86996 &  85209 & 84306 &  82531 &  83184 &  1.2179 &  82430&         &  -754 \\
$5s^25p^57s$ & $^2$[3/2]$^o$ & 2 &  80504 &  89101 &  86967 & 86367 &  84254 &  85042 &  1.4910 &  85189&         &   147 \\
             &               & 1 &  81064 &  89742 &  87648 & 86951 &  84880 &  85655 &  0.9759 &  85440&         &  -215 \\
$5s^25p^57p$ & $^2$[1/2]     & 1 &  83048 &  91935 &  89859 & 89051 &  87001 &  87773 &  1.7930 &  87927&  1.7272 &   154 \\
             &               & 0 &  84221 &  92910 &  90834 & 90110 &  88057 &  88825 &  0      &  88842&         &    17 \\
$5s^25p^57p$ & $^2$[5/2]     & 2 &  83464 &  92252 &  90160 & 89397 &  87330 &  88104 &  1.1107 &  88352&  1.1276 &   248 \\
             &               & 3 &  83558 &  92350 &  90262 & 89494 &  87429 &  88204 &  1.3333 &  88469&  1.330  &   265 \\
$5s^25p^56p$ & $^2$[3/2]     & 1 &  83700 &  92521 &  90429 & 89662 &  87594 &  88369 &  1.0216 &  88379&  0.7925 &    10 \\
             &               & 2 &  84883 &  94445 &  92365 & 91398 &  89352 &  90122 &  1.1497 &  89162&  1.190  &  -960 \\
$5s^25p^56d$ & $^2$[1/2]$^o$ & 0 &  83637 &  92398 &  90367 & 89562 &  87557 &  88309 &  0      &  88491&         &   182 \\
             &               & 1 &  83728 &  92489 &  90466 & 89654 &  87658 &  88405 &  1.3430 &  88550&         &   145 \\
$5s^25p^57p$ & $^2$[3/2]     & 2 &  83832 &  92610 &  90519 & 89759 &  87693 &  88466 &  1.3843 &  88687&  1.3520 &   221 \\
             &               & 1 &  84392 &  94128 &  92055 & 90956 &  88909 &  89680 &  0.6345 &  88745&  0.9039 &  -935 \\
$5s^25p^56d$ & $^2$[3/2]$^o$ & 2 &  83947 &  92702 &  90695 & 89870 &  87892 &  88632 &  1.3165 &  88708&         &    76 \\
             &               & 1 &  86666 &  95327 &  93442 & 92542 &  90699 &  91378 &  0.6980 &  90032&         & -1346 \\
$5s^25p^56d$ & $^2$[7/2]$^o$ & 4 &  84071 &  92767 &  90743 & 89952 &  87956 &  88701 &  1.2500 &  88912&         &   211 \\
             &               & 3 &  84273 &  92937 &  90939 & 90134 &  88167 &  88900 &  1.0926 &  89025&         &   125 \\
$5s^25p^56d$ & $^2$[5/2]$^o$ & 2 &  84610 &  93238 &  91272 & 90449 &  88514 &  89232 &  0.9548 &  89243&         &    11 \\
             &               & 3 &  84900 &  93476 &  91533 & 90700 &  88790 &  89496 &  1.2085 &  89535&         &    39 \\
    \end{tabular}
  \end{ruledtabular}
\end{table*}

Results for neutral xenon are presented in Table~\ref{Xe} while results for
six positive ions from Xe~VII to Xe~II are presented in
Tables~\ref{XeVII}, \ref{XeVI}, \ref{XeV}, \ref{XeIV},\ref{XeIII} and \ref{XeII}.  
For neutral xenon (Table~\ref{Xe}) we study in detail the role of core-valence
correlations by including them in different approximations. 
The basis for valence states is kept the same in all cases (see previous
section for the description of the basis). 
The approximations are
\begin{enumerate}
\item First, we present the results of the standard CI method, with no
core-valence correlations (the ``CI'' column of Table~\ref{Xe}). Accuracy for
the energies as compared to experimental values are not very good. However the
difference does not exceed 10\% which is sufficiently good for many
applications. This is in spite of the fact that calculations for neutral xenon
were done with atomic core corresponding to highly ionized Xe~IX. This is
another confirmation that change in the core potential $V_{core}$ from Xe~I to
Xe~IX is very small.
\item Second-order $\hat \Sigma_1^{(2)}$ is added to the effective Hamiltonian
(the ``$\hat \Sigma_1^{(2)}$'' column of Table~\ref{Xe}). The results are
significantly closer to the experiment but the correction is too large. This
is similar to what usually takes place with the second-order correlation
correction for atoms with one external electron.
\item Second-order $\hat \Sigma_2^{(2)}$ is also added (the ``$\hat
  \Sigma_1^{(2)} \& \hat \Sigma_2^{(2)}$'' column). As one can see $\hat
  \Sigma_2^{(2)}$ acts in opposite direction to $\hat \Sigma_1^{(2)}$ and the
  results are even closer to the experiment.
\item Higher orders are included in $\hat \Sigma_1$ while $\hat \Sigma_2$ is
  not included at all (the ``$\hat \Sigma_1^{(\infty)}$'' column). The effect
  of higher orders in $\hat \Sigma_1$ is numerically close to the
 effect of $\hat
  \Sigma_2$ as is evident from the comparison with previous column.
 This coincidence is accidental.
\item Higher orders are included in $\hat \Sigma_1$ while $\hat \Sigma_2$ is
  included in second order (the ``$\hat \Sigma_1^{(\infty)} \& \hat
  \Sigma_2^{(2)}$'' column). The results are improved but for many states the
  correction is too large.
\item Higher orders are included in both $\hat \Sigma_1$ and $\hat \Sigma_2$ 
(the ``$\hat \Sigma^{(\infty)}$'' column). This is the most complete
calculation we
have in present work. 
Here we also included calculated values of Land\'{e}'s $g$-factors.
The $g$-factors are very useful for identification of the states,
especially for atoms with dense spectrum where calculations do not always
reproduce the correct order of the levels.
\end{enumerate}
The last column of Table~\ref{Xe} presents the difference between experimental
and calculated energies where calculated energies correspond to the most
complete calculation ($\hat \Sigma^{(\infty)}$): $\Delta = E_{expt} - E_{calc}$. This
difference does not exceed 2\% and should mostly be atributed to
incompleteness of the basis. Indeed, it is hard to expect that the basis
consisting of only 15 single-electron states (from one to three in each partial
wave from $l=0$ to $l=3$) to be complete. Test calculations show that adding
more states to the basis do have some effect on the energies of the
states. The effect is larger for higher states. For example, it is hard to
expect any reasonable accuracy for the states of the $5s^25p^56d$
configuration without having the $6d$ state in the basis. But adding the $6d$
state to the basis also have some effect on the lower $5s^25p^55d$
configuration. The detailed study of the ways to saturate the basis goes
beyond the scope of the present work.

Tables~\ref{XeVII}, \ref{XeVI}, \ref{XeV}, \ref{XeIV},\ref{XeIII} and \ref{XeII}.  
present our results for
xenon positive ions from Xe~VII to Xe~II. Only results obtained in the
``best'' approximation ($\hat \Sigma^{(\infty)}$) are included. Calculations
for the ions start from point 4 in the scheme presented in the beginning of
this section. This is because first 3 points are exactly the same as for
neutral xenon. Note that one of the most time consuming steps, calculation of 
$\hat \Sigma_1$, doesn't need to be repeated. Basis states for valence
 electrons
used in the CI calculations are described in previous section (see
Table~\ref{basis}). We use shorter basis for the ions because we calculate
only lowest states. To go up in the spectrum we would need to extend the basis
similar to what is done for Xe~I. The analysis of the data in 
Tables~\ref{XeVII}, \ref{XeVI}, \ref{XeV}, \ref{XeIV},\ref{XeIII} and \ref{XeII}.  
show that the accuracy is generally very good in spite of very short basis.

For the Xe~III ion we also included calculations which use the basis states of
the Xe~IV ion (column $E(N-1)$ of Table~\ref{XeIII}). The purpose of these
calculations will be explained in the {\em negative ions} section below.

\begin{table}
  \caption{Ground state removal energy (a.u.), excitation energies
(cm$^{-1}$) and $g$-factors of lowest states of Xe~VII ; 
comparison with experiment}
  \label{XeVII}
  \begin{ruledtabular}
    \begin{tabular}{l l c r r l r}
      \multicolumn{2}{c}{State} & $J$ & \multicolumn{1}{c}{Expt~\cite{NIST}} &
      \multicolumn{2}{c}{Calculations} & \\ 
 & & & \multicolumn{1}{c}{$E$} & \multicolumn{1}{c}{$E$} & 
 \multicolumn{1}{c}{$g$} &
 \multicolumn{1}{c}{$\Delta$} \\
 \hline
$5s^2$ & $^1$S     & 0 &   -7.26 &         &   -7.27  & \\
\hline
$5s5p$ & $^3$P$^o$ & 0 &   96141 &   94889 &   0      &   1252 \\
       &           & 1 &  100451 &   99394 &   1.4846 &   1057 \\
       &           & 2 &  113676 &  112598 &   1.5000 &   1078 \\
$5s5p$ & $^1P$$^o$ & 1 &  143259 &  146337 &   1.0153 &  -3078 \\
$5p^2$ & $^3$P     & 0 &  223673 &  224343 &   0      &   -670 \\
       &           & 1 &  234685 &  235008 &   1.5000 &   -323 \\
       &           & 2 &  251853 &  252607 &   1.3027 &   -754 \\
$5p^2$ &   $^1$D   & 2 &  236100 &  237129 &   1.1962 &  -1029 \\
$5p^2$ &   $^1$S   & 0 &  273208 &  281328 &   0      &  -8120 \\
$5s5d$ &   $^3$D   & 1 &  287772 &  291855 &   0.5000 &  -4083 \\
       &           & 2 &  288712 &  292896 &   1.1663 &  -4184 \\
       &           & 3 &  290340 &  294591 &   1.3333 &  -4251 \\
$5s5d$ &   $^1$D   & 2 &  307542 &  317647 &   1.0015 & -10105 \\
$5s6s$ & $^3$S     & 1 &  354833 &  358686 &   2.0000 &  -3853 \\
$5s6s$ & $^1$S     & 0 &  361671 &  364853 &   0      &  -3182 \\
$5p5d$ & $^3$F$^o$ & 2 &  393792 &  398186 &   0.7405 &  -4394 \\
       &           & 3 &  401413 &  406187 &   1.0990 &  -4774 \\
       &           & 4 &  412567 &  417863 &   1.2500 &  -5296 \\
    \end{tabular}
  \end{ruledtabular}
\end{table}

\begin{table}
  \caption{Ground state removal energy (a.u.), excitation energies
(cm$^{-1}$) and $g$-factors of lowest states of Xe~VI ; 
comparison with experiment}
  \label{XeVI}
  \begin{ruledtabular}
    \begin{tabular}{l l c r r l r}
      \multicolumn{2}{c}{State} & $J$ & \multicolumn{1}{c}{Expt~\cite{NIST}} &
      \multicolumn{2}{c}{Calculations} & \\ 
 & & & \multicolumn{1}{c}{$E$} & \multicolumn{1}{c}{$E$} & 
 \multicolumn{1}{c}{$g$} &
 \multicolumn{1}{c}{$\Delta$} \\
 \hline
$5s^25p$ &  $^2$P$^o$ & 1/2 &  -9.71 &    -9.72 &  0.6667 &  \\
\hline
         &            & 3/2 &  15599 &    15590 &  1.3333 &      9  \\
$5s5p^2$ &  $^4$P     & 1/2 &  92586 &    90191 &  2.6301 &   2395  \\
         &            & 3/2 & 100378 &    97787 &  1.7247 &   2591  \\
         &            & 5/2 & 107205 &   105036 &  1.5629 &   2169  \\
$5s5p^2$ &  $^2$D     & 3/2 & 124870 &   125900 &  0.8243 &  -1030  \\
         &            & 5/2 & 129230 &   129897 &  1.2366 &   -667  \\
$5s5p^2$ &  $^2$P     & 1/2 & 141837 &   145429 &  1.1878 &  -3592  \\
         &            & 3/2 & 159112 &   162903 &  1.3119 &  -3791  \\
$5s5p^2$ &  $^2$S     & 1/2 & 157996 &   161647 &  1.5155 &  -3651  \\
$5s^25d$ &  $^2$D     & 3/2 & 180250 &   186188 &  0.8058 &  -5938  \\
         &            & 5/2 & 182308 &   188093 &  1.2004 &  -5785  \\
$5s^26s$ &  $^2$S     & 1/2 & 223478 &   224641 &  1.9998 &  -1163  \\
$5p^3$   &            & 3/2 & 232586 &   232997 &  1.3377 &   -411  \\
    \end{tabular}
  \end{ruledtabular}
\end{table}

\begin{table}
  \caption{Ground state removal energy (a.u.), excitation energies
(cm$^{-1}$) and $g$-factors of lowest states of Xe~V; 
comparison with experiment}
  \label{XeV}
  \begin{ruledtabular}
    \begin{tabular}{l l c r r l r}
      \multicolumn{2}{c}{State} & $J$ & \multicolumn{1}{c}{Expt~\cite{NIST}} &
      \multicolumn{2}{c}{Calculations} & \\ 
 & & & \multicolumn{1}{c}{$E$} & \multicolumn{1}{c}{$E$} & 
 \multicolumn{1}{c}{$g$} &
 \multicolumn{1}{c}{$\Delta$} \\
 \hline
$5s^25p^2$ &  $^3$P      & 0 &  -11.7 &   -11.72 &         & \\
\hline
           &             & 1 &   9292 &     8969 &  1.5000 &    323 \\
           &             & 2 &  14127 &    14643 &  1.3744 &   -516 \\
$5s^25p^2$ &  $^1$D      & 2 &  28412 &    30169 &  1.1256 &  -1757 \\
$5s^25p^2$ &  $^1$S      & 0 &  44470 &    47061 &  0      &  -2591 \\
$5s5p^3$   &  $^5$S$^o$  & 2 &  92183 &    88033 &  1.9744 &   4150 \\
$5s5p^3$   &  $^3$D$^o$  & 1 & 115286 &   115554 &  0.6192 &   -268 \\
           &             & 2 & 116097 &   116202 &  1.2256 &   -105 \\
           &             & 3 & 119919 &   120152 &  1.3329 &   -233 \\
$5s5p^3$   &  $^3$P$^o$  & 0 & 133408 &   134320 &  0      &   -912 \\
           &             & 1 & 134575 &   135493 &  1.4078 &   -918 \\
           &             & 2 & 134703 &   135579 &  1.3152 &   -876 \\
$5s5p^3$   &  $^1$D$^o$  & 2 & 145807 &   147030 &  1.1261 &  -1223 \\
$5s5p^3$   &  $^3$S$^o$  & 1 & 155518 &   160672 &  1.7362 &  -5154 \\
$5s^25p5d$ &  $^3$F$^o$  & 2 & 156507 &   159419 &  0.7036 &  -2912 \\
           &             & 3 & 160630 &   163534 &  1.0901 &  -2904 \\
           &             & 4 & 169799 &   172418 &  1.2500 &  -2619 \\
$5s5p^3$   &  $^1$P$^o$  & 1 & 169673 &   175704 &  1.1706 &  -6031 \\
    \end{tabular}
  \end{ruledtabular}
\end{table}

\begin{table}
  \caption{Ground state removal energy (a.u.), excitation energies
(cm$^{-1}$) and $g$-factors of lowest states of Xe~IV}
  \label{XeIV}
  \begin{ruledtabular}
    \begin{tabular}{l l c r r l r}
      \multicolumn{2}{c}{State} & $J$ & \multicolumn{1}{c}{Expt~\cite{NIST}} &
      \multicolumn{2}{c}{Calculations} & \\ 
 & & & \multicolumn{1}{c}{$E$} & \multicolumn{1}{c}{$E$} & 
 \multicolumn{1}{c}{$g$} &
 \multicolumn{1}{c}{$\Delta$} \\
 \hline
$5s^25p^3$   & $^4$S$^o$ & 3/2 &  -13.2 & -13.27 &   1.8987 & \\
\hline
$5s^25p^3$   & $^2$D$^o$ & 3/2 &  13267 &  14619 &   0.9778 &  -1352 \\
             &           & 5/2 &  17511 &  18938 &   1.2000 &  -1427 \\
$5s^25p^3$   & $^2$P$^o$ & 1/2 &  28036 &  30149 &   0.6667 &  -2113 \\
             &           & 3/2 &  35650 &  37446 &   1.2569 &  -1796 \\
$5s5p^4$     & $^4$P     & 5/2 &  99664 &  99466 &   1.5814 &    198 \\
             &           & 3/2 & 106923 & 106710 &   1.7055 &    213 \\
             &           & 1/2 & 109254 & 109169 &   2.6286 &     85 \\
$5s5p^4$     & $^2$D     & 3/2 & 121929 & 124529 &   0.8925 &  -2600 \\
             &           & 5/2 & 125475 & 128117 &   1.2153 &  -2642 \\
$5s^25p^25d$ & $^2$P     & 3/2 & 133027 & 135880 &   0.8912 &  -2853 \\
             &           & 1/2 & 136796 & 139997 &   0.7788 &  -3201 \\
$5s^25p^25d$ & $^4$F     & 3/2 & 134981 & 137617 &   0.8501 &  -2636 \\
             &           & 5/2 & 136496 & 139103 &   1.1064 &  -2607 \\
             &           & 7/2 & 141625 & 144013 &   1.2649 &  -2388 \\
             &           & 9/2 & 145991 & 148958 &   1.3088 &  -2967 \\
$5s^25p^25d$ & $^2$F     & 5/2 & 141824 & 145598 &   0.9889 &  -3774 \\
             &           & 7/2 & 145011 & 148526 &   1.2658 &  -3515 \\
$5s^25p^25d$ & $^4$D     & 1/2 & 145106 & 147933 &   0.4037 &  -2827 \\
             &           & 3/2 & 146207 & 148762 &   1.1487 &  -2555 \\
             &           & 5/2 & 148685 & 151840 &   1.1799 &  -3155 \\
             &           & 7/2 & 155864 & 159785 &   1.2361 &  -3921 \\
$5s5p^4$     & $^2$S     & 1/2 & 150737 & 154437 &   1.5219 &  -3700 \\
$5s^25p^26s$ & $^4$P     & 1/2 & 157205 & 161777 &   2.3294 &  -4572 \\
             &           & 3/2 & 165280 & 167775 &   1.6017 &  -2495 \\
             &           & 5/2 & 170490 & 174875 &   1.4873 &  -4385 \\
$5s^25p^25d$ & $^4$P     & 5/2 & 159643 & 165187 &   1.5539 &  -5544 \\
             &           & 3/2 & 161435 & 169550 &   1.6737 &  -8115 \\
             &           & 1/2 & 162867 & 169085 &   2.4382 &  -6218 \\
    \end{tabular}
  \end{ruledtabular}
\end{table}

\begin{table}
  \caption{Ground state removal energy (a.u.), excitation energies
(cm$^{-1}$) and $g$-factors of lowest states of Xe~III}
  \label{XeIII}
  \begin{ruledtabular}
    \begin{tabular}{l l c r l r l r r}
      \multicolumn{2}{c}{State} & $J$ & \multicolumn{2}{c}{Expt~\cite{NIST}} &
      \multicolumn{4}{c}{Calculations}  \\ 
 & & & \multicolumn{1}{c}{$E$} & \multicolumn{1}{c}{$g$} & 
\multicolumn{1}{c}{$E$} & \multicolumn{1}{c}{$g$} & \multicolumn{1}{c}{$\Delta$} &
\multicolumn{1}{c}{$E(N-1)$} \\
\hline
$5s^25p^4$   & $^3$P      & 2 & -14.38  &      & -14.36 &  1.4523 &       & -14.38\\
\hline
             &            & 0 &   8130  &      &   8313 &  0      &  -183 &  8319 \\
             &            & 1 &   9794  &      &   9638 &  1.5000 &   156 & 9528\\
$5s^25p^4$   & $^1$D      & 2 &  17099  &      &  19086 &  1.0477 & -1987 & 17879\\
$5s^25p^4$   & $^1$S      & 0 &  36103  &      &  37280 &  0      & -1177 & 37392\\
$5s5p^5$     & $^3$P$^o$  & 2 &  98262  &      &  98847 &  1.4986 &  -585 & 98729\\
             &            & 1 & 103568  &      & 104334 &  1.4553 &  -766 & 104369\\
             &            & 0 & 108334  &      & 108562 &  0      &  -228 & 109016\\
$5s^25p^35d$ & $^5$D$^o$  & 3 & 111605  &      & 110836 &  1.4702 &   769 & 113883\\
             &            & 2 & 111856  &      & 111066 &  1.4623 &   790 & 114075\\
             &            & 4 & 112272  &      & 111366 &  1.4813 &   906 & 114347\\
             &            & 1 & 112450  &      & 111506 &  1.4882 &   944 & 114360\\
             &            & 0 & 112694  &      & 112142 &  0      &   552 & 114544\\
$5s^25p^35d$ & $^3$D$^o$  & 2 & 117240  &      & 118575 &  1.8400 & -1335 & 122839\\
             &            & 3 & 121230  &      & 122482 &  1.3114 & -1252 & 126365\\
             &            & 1 & 121923  &      & 123251 &  0.6428 & -1328 & 127257\\
$5s^25p^35d$ & $^1$P$^o$  & 1 & 119026  &      & 120856 &  1.0670 & -1830 & 122930\\
$5s^25p^36s$ & $^5$S$^o$  & 2 & 121476  & 1.95 & 119002 &  1.1889 &  2474 & 125150\\
$5s^25p^35d$ & $^3$F$^o$  & 2 & 124691  &      & 127328 &  0.8030 & -2637 & 130640\\
             &            & 3 & 126120  &      & 129045 &  1.0994 & -2925 & 132337\\
             &            & 4 & 130174  &      & 133970 &  1.1386 & -3796 & 138195\\
$5s^25p^36s$ & $^3$S$^o$  & 1 & 125617  & 1.77 & 124152 &  1.7102 &  1465 & 132504\\
$5s^25p^35d$ & $^3$G$^o$  & 4 & 127782  &      & 131307 &  1.1977 & -3525 & 134519\\
             &            & 3 & 128349  &      & 132922 &  0.8245 & -4573 & 137183\\
             &            & 5 & 132160  &      & 136422 &  1.2000 & -4262 & 140435\\
    \end{tabular}
  \end{ruledtabular}
\end{table}

\begin{table}
  \caption{Ground state removal energy (a.u.), excitation energies
(cm$^{-1}$) and $g$-factors of lowest states of Xe~II}
  \label{XeII}
  \begin{ruledtabular}
    \begin{tabular}{l l c r l r l r}
      \multicolumn{2}{c}{State} & $J$ & \multicolumn{2}{c}{Expt~\cite{NIST}} &
      \multicolumn{2}{c}{Calculations} & \\ 
 & & & \multicolumn{1}{c}{$E$} & \multicolumn{1}{c}{$g$} & 
\multicolumn{1}{c}{$E$} & \multicolumn{1}{c}{$g$}  & \multicolumn{1}{c}{$\Delta$} \\
\hline
$5s^25p^5$    & $^2$P$^o$  &   3/2 &  -15.16   &       &  -15.09  &  1.3333 &\\
\hline
              &            &   1/2 &   10537   &       &   10763  &  0.6667 &    -226\\
$5s5p^6$      & $^2$S      &   1/2 &   90874   &  2.02 &   91700  &  2.0423 &    -826\\
$5s^25p^46s$  &  [2]       &   5/2 &   93068   &  1.56 &   91729  &  1.5639 &    1339\\
              &            &   3/2 &   95064   &  1.38 &   94188  &  1.4084 &     876\\
$5s^25p^45d$  &  [2]       &   5/2 &   95397   &  1.36 &   95802  &  1.3473 &    -405\\
              &            &   3/2 &   96033   &  1.18 &   96534  &  1.1847 &    -501\\
$5s^25p^45d$  &  [3]       &   7/2 &   95438   &  1.42 &   95783  &  1.3940 &    -345\\
              &            &   5/2 &  106475   &       &  108559  &  1.0537 &   -2084\\
$5s^25p^45d$  &  [1]       &   1/2 &   96858   &  0.50 &   97388  &  0.5457 &    -530\\
              &            &   3/2 &  105313   &  1.15 &  107286  &  1.0869 &   -1973\\
$5s^25p^45d$  &  [4]       &   9/2 &   99405   &  1.31 &  100848  &  1.3093 &   -1443\\
              &            &   7/2 &  101536   &  1.11 &  103581  &  1.1524 &   -2045\\
$5s^25p^46s$  &  [0]       &   1/2 &  101157   &  2.43 &  100700  &  2.3677 &     457\\
$5s^25p^46s$  &  [1]       &   3/2 &  102799   &  1.59 &  101988  &  1.5811 &     811\\
              &            &   1/2 &  106906   &  1.79 &  108148  &  1.9490 &   -1242\\
$5s^25p^45d$  &  [1]       &   1/2 &  104250   &  0.56 &  104264  &  0.6408 &     -14\\
              &            &   3/2 &  107904   &  1.20 &  109217  &  1.3762 &   -1313\\
$5s^25p^45d$  &  [0]       &   1/2 &  105948   &  1.36 &  107566  &  1.1311 &   -1618\\
$5s^25p^46p$  &  [2]$^o$   &   3/2 &  111792   &  1.61 &  112248  &  1.6084 &    -456\\
              &            &   5/2 &  111959   &  1.47 &  112286  &  1.4934 &    -327\\
$5s^25p^46p$  &  [3]$^o$   &   5/2 &  113512   &  1.28 &  114041  &  1.2350 &    -529\\
              &            &   7/2 &  113705   &  1.40 &  114266  &  1.3984 &    -561\\
$5s^25p^46p$  &  [1]$^o$   &   1/2 &  113673   &  1.50 &  114350  &  1.5358 &    -677\\
              &            &   3/2 &  116783   &  1.37 &  117621  &  1.3609 &    -838\\
    \end{tabular}
  \end{ruledtabular}
\end{table}

\section{some special cases}

\subsection{Highly excited states}

One of the additional advantages of the use of $V^N$ basis for valence states
is the possibility to study highly excited states with a very short basis.
To get to a highly excited state with an universal basis like B-splines one 
has to calculate all states of the same parity and total momentum $J$ 
which are below of the state of interest. Also, the completeness of the
basis deteriorates rapidly while going  higher in the spectrum.
 $V^N$ basis is free from these problems. To calculate highly excited states
of a particular configuration it is sufficient to include into single-electron
basis for valence states only states which correspond to this configuration. 
For example, the states of the $5s^25p^58s$ of Xe~I can be calculated with good
accuracy with only four states in the basis: $5s$, $5p_{1/2}$, $5p_{3/2}$ and
$8s$ (see Table~\ref{Ryd}). There are many lower states of same parity and 
total momentum $J$ but we can easily get rid of them by not including
corresponding single-electron states into the basis.  

\begin{table}
  \caption{Energies (cm$^{-1}$) and $g$-factors of the $5s^25p^58s$ 
configuration of Xe~I}
  \label{Ryd}
  \begin{ruledtabular}
    \begin{tabular}{l c r l r l}
      \multicolumn{1}{c}{State} & $J$ & \multicolumn{2}{c}{Expt.~\cite{NIST}} &
\multicolumn{2}{c}{Calculations} \\
                &   & \multicolumn{1}{c}{$E$} & \multicolumn{1}{c}{$g$} & 
                      \multicolumn{1}{c}{$E$} & \multicolumn{1}{c}{$g$} \\
\hline
$8s$ [3/2]$^o$  & 2 &  90805 &  1.465 &  92288 &  1.5000 \\
                & 1 &  90933 &  1.182 &  92414 &  1.1700 \\

$8s'$ [1/2]$^o$ & 0 &        &        & 104063 &  0      \\
                & 1 & 101426 &        & 104118 &  1.3300 \\
    \end{tabular}
  \end{ruledtabular}
\end{table}

\subsection{Negative ions}

An interesting question is whether method presented in this paper can be used
to calculated states of a negative ion. At first glance the answer is {\em no} 
because we use  $V^N$ states for the basis and negative ions are not bound
in the  $V^N$ approximation. However, we may consider the following question:
what is going to happen if we add one more electron to the CI calculations
for a neutral atom, when basis corresponds to the neutral atom?

For atoms like xenon, which don't form negative ions, it makes more sense
to consider more general question: can a basis calculated for a system of
$N-1$ valence electrons be used to calculate many electron states of a
system of $N$ electrons? This can be easily checked. Take, for example,
the Xe~IV ion and add one more electron in the CI calculations 
to get the states of Xe~III. We've done this without adding any new states
into the basis. Results are presented in the last column of
Table~\ref{XeIII}. We can see that the results for the ion with the basis of
the other ion are almost as good as with its own basis. Accuracy is a bit
lower which is a natural consequence of the worsening of the basis. Adding more
states to the basis would most certainly improve the results.

This findings are not
very surprising since we know that any basis set can be used in the CI
calculations. For example, in Ref.~\cite{vn} calculations for neutral Kr
were performed with the basis corresponding to Kr~IX! The only question is 
how many states we need to include to get reasonable results. It
turns out that at least in the case of just one more electron there is
no need to greatly increase the basis. This means that we can also
calculate states of negative ions by using basis states of a neutral
atom. 

\section{Conclusion}

In this paper we present a method of calculation for many-electron atoms with
open shells which is both accurate and very efficient. The method is based on
the so called $V^{N-M}$ approximation in which calculations start from the
highly charged ion with all valence electrons removed. High accuracy is
achieved by inclusion of core-valence correlations by means of MBPT. 
Dominating chains of higher order diagrams are included in all orders. 
High efficiency of the method is mostly due to the compact  $V^N$ basis set for
the states of valence electrons. The method is expected to work well
for atoms in which valence electrons form a separate shell (defined by the
principal quantum number). This is usually the case if valence electrons in
the atomic ground state occupy $s$ and/or $p$ states. This covers roughly half
of the periodic table of elements. Calculations for xenon and its ions
illustrate the use of the method.

\section*{Acknowledgments}

The work was funded in part by the Australian Research Council.

\appendix

\section{Efficient way of calculating $\hat \Sigma$}

The correlation correction operator $\hat \Sigma_1$ is defined in such are
way that its overage value over a wave function of a valence electron is
the correlation correction to the energy of this electron:
\begin{equation}
  \delta \epsilon_v = \langle v | \hat \Sigma_1 | v \rangle.
  \label{sigma1def}
\end{equation}
We use the following form of the single-electron wave function
\begin{equation}
  \psi(r)_{njlm}=\frac{1}{r}\left(\begin {array}{c} 
    f_{n}(r)\Omega(\mathbf{n})_{\mathit{jlm}}  \\ 
    i\alpha g_{n}(r)  \tilde{ \Omega}(\mathbf{n})_{\mathit{jlm}} 
  \end{array} \right).
  \label{psi}
\end{equation}
Then the expression (\ref{sigma1def}) becomes
\begin{eqnarray}
  \delta \epsilon_v &=&    \int \int f_v(r) \Sigma_{ff}(r,r') f_v(r')drdr'  \\
           +    &\alpha^2& \int \int f_v(r) \Sigma_{fg}(r,r') g_v(r')drdr' \nonumber \\
	   +    &\alpha^2& \int \int g_v(r) \Sigma_{gf}(r,r') f_v(r')drdr' \nonumber \\
	   +    &\alpha^4& \int \int g_v(r) \Sigma_{gg}(r,r') g_v(r')drdr'. \nonumber
  \label{sigmaff}
\end{eqnarray}
Note factors $\alpha^2$ and $\alpha^4$ in all terms except the first one. These
factors make corresponding contributions to be very small. Therefore we don't
usually include them. Only $\Sigma_{ff}$ will be considered in this appendix and
we will omit the indexes.

\subsection{Second-order $\hat \Sigma$}

Good efficiency in calculating of $\hat \Sigma$ is achieved by dividing the
calculations into two steps:
\begin{enumerate}
\item First, all relevant Coulomb $Y$ functions are calculated and stored on disk.
\item Then, $\Sigma$ is calculated using stored $Y$-functions.
\end{enumerate}
The Coulomb $Y$-function is defined as
\begin{eqnarray}
 & Y_{knm}(r) = \int \frac{r_<^k}{r_>^{k+1}}(f_n(r')f_m(r') + \nonumber \\
 & \alpha^2 g_n(r')g_m(r'))dr',
\label{Yknm}
\end{eqnarray}
where $r_< = {\rm min}(r,r')$ and $r_> = {\rm max}(r,r')$.
We will also need a $\rho$ function:
\begin{equation}
  \rho_{jl}(r) = f_j(r)f_l(r)+\alpha^2 g_j(r)g_l(r)).
\label{rho}
\end{equation}
Our typical coordinate grid consists of about 1000 points. Usually all of them are
used to calculate $Y$-functions (\ref{Yknm}). However, there is no need to keep
all points for the $Y$ and $\rho$-functions for consequent calculations.
It turns out that very little lose of accuracy is caused by the use of a
subset of points defined as every 4th point in the interval
\[ 1/Z \leq r \leq R_{core}. \]
By cutting off the point on short and large distances and using only every 4th
point in between we reduce the number of points by an order of magnitude.
Then, Coulomb integrals are calculated in an extremely efficient way
\begin{equation}
  q_k(jlmn) = \sum_{i=1}^{\mu}\rho_{jl}(r_i)Y_{kmn}(r_i)w_i.
\label{qk}
\end{equation}
Here $\mu \approx 100$ is number of points on the sub-grid and $w_i$ are
weight coefficients corresponding to a particular method of numerical
integration. Note that only one of two integrations for Coulomb
integrals is done on a reduced sub-grid. First integration (\ref{Yknm})
is done with the use of all points. 

An expressions for $\hat \Sigma_1^{(2)}$  via $Y$-functions is
\begin{eqnarray}
  &\Sigma_1(r,r') = \nonumber \\ 
  & \sum_{amnk} c_1(kvamn) \times \nonumber \\ 
  & \frac{f_n(r)Y_{kam}(r)Y_{kam}(r')f_n(r')}
  {\epsilon_v+\epsilon_a-\epsilon_m-\epsilon_n} \label{s1} \\
  & - \sum_{amnk_1k_2} c_2(k_1k_2vamn) \times \nonumber \\ 
  & \frac{f_n(r)Y_{k_1am}(r)Y_{k_2an}(r')f_m(r')}
  {\epsilon_v+\epsilon_a-\epsilon_m-\epsilon_n}  \label{s2} \\
  & - \sum_{amnk} c_3(kvabm) \times \nonumber \\ 
  & \frac{f_b(r)Y_{kam}(r)Y_{kam}(r')f_b(r')}
  {\epsilon_a+\epsilon_b-\epsilon_v-\epsilon_m} \label{s3} \\
  & + \sum_{amnk_1k_2} c_4(k_1k_2vabm) \times \nonumber \\ 
  & \frac{f_a(r)Y_{k_1bm}(r)Y_{k_2am}(r')f_b(r')}
  {\epsilon_a+\epsilon_b-\epsilon_v-\epsilon_m} \label{s4}
\label{diag1}
\end{eqnarray}
Here $c_1$, $c_2$, $c_3$, $c_4$ are angular coefficients. Expressions
for them can be found elsewhere~\cite{CI+MBPT}.
Formulas (\ref{s1}),(\ref{s2}),(\ref{s3}),(\ref{s4}), 
correspond to diagrams 1,2,3,4 on Fig.~\ref{sigma1-fig}.
$\hat \Sigma_1$ is a matrix of size $\mu \approx 100$ in
coordinate space. Matrix elements of $\hat \Sigma_1$ are calculated by
\begin{eqnarray}
  & \langle v | \hat \Sigma_1 | w \rangle = \nonumber \\
  & \sum_{i=1,j=1}^{\mu} f_v(r_i) \Sigma_1(r_i,r_j) f_w(r_j)w_iw_j.
\label{sigma1me}
\end{eqnarray}
Note that we use a two-step procedure to calculate matrix elements
of $\hat \Sigma_1$. First, $\hat \Sigma_1$ matrix which is independent 
on valence functions is calculated, then matrix elements of $\hat \Sigma_1$ 
are calculated. To use the same approach for $\hat \Sigma_2$ is impractical.
As can be seen from Fig.~\ref{sigma2-fig} to make $\hat \Sigma_2$ 
independent on valence states one would have to make matrices of 
dimensions 2, 3 and 4. Therefore we just calculate matrix elements of
$\hat \Sigma_2$ via Coulomb integrals. Corresponding expressions can 
be found in Ref.~\cite{CI+MBPT}. Coulomb integrals are calculated as in (\ref{qk}).

\subsection{All-order $\Sigma$}

We use Feynman diagram technique to include higher-order correlations into
direct part of $\hat \Sigma_1$ (diagrams 1 and 3 on Fig.~\ref{sigma1-fig}).
Corresponding expression is~\cite{CsPNC}
\begin{equation}
  \Sigma(\epsilon, r_i,r_j)= \sum_{nm} \int \frac{d\omega}{2\pi}G_{ij}(\epsilon+\omega)
  Q_{im}\tilde \Pi_{mn}(\omega)Q_{nj}.
\label{sigma-all}
\end{equation}
Here $\tilde \Pi$ is ``screened polarization operator'' 
\[ \tilde \Pi = \Pi[1-Q\Pi]^{-1}, \] $\Pi$ is polarization operator 
\[ \Pi(\omega) = \sum_a \psi_a [G(\epsilon_a+\omega)+G(\epsilon_a-\omega)]\psi_a, \] 
$G$ is Green function 
\[ (\hat h_1 - \epsilon)G(r,r') = -\delta(r-r'), \] 
and $Q$ is Coulomb interaction \[ Q_{ij} = e^2/(r_i-r_j). \]
The details of calculation of $\hat \Sigma_1^{(\infty)}$ can be found 
elsewhere~\cite{DSF89,CsPNC}. Here we only want to mention that
operators $\tilde \Pi$, $\Pi$, $G$ and $Q$ are matrices of size 
$\mu \approx 100$ in coordinate space. Therefore calculation of
$\hat \Sigma_1^{(\infty)}$ involves manipulation of matrices of
relatively small size. If we also recall that $\hat \Sigma_1$
does not depend on valence states and needs to be calculated only 
once then the efficiency of its calculation is quiet satisfactory.

Higher-order correlations in exchange diagrams for $\hat \Sigma_1$ 
(diagrams 2 and 4 on Fig.~\ref{sigma1-fig}) and for all diagrams
for $\hat \Sigma_2$ are included via screening factors as explained
in the text.


\begin{thebibliography}{999}

\frenchspacing

\bibitem{Ginges} J. S. M. Ginges and V. V. Flambaum, Physics Reports,
{\bf 397} 63 (2004).

\bibitem{alpha} J. K. Webb, V. V. Flambaum, C. W. Churchill,
M. J. Drinkwater, and J. D. Barrow, Phys. Rev. Lett. {\bf 82}, 884 (1999);
J. K. Webb, M. T. Murphy, V. V. Flambaum, V. A. Dzuba, J. D. Barrow,
C. W. Churchill, J. X. Prochaska, and A. M. Wolfe,
Phys. Rev. Lett. {\bf 87}, 091301 (2001);
M. T. Murphy, J. K. Webb, V. V. Flambaum, V. A. Dzuba, 
C. W. Churchill, J. X. Prochaska, J. D. Barrow, and A. M. Wolfe,
Mon. Not. R. Astron. Soc. {\bf 327}, 1208 (2001);
\bibitem{clock} E. J. Angstmann, V. A. Dzuba, V. V. Flambaum, 
      Phys. Rev. A, {\bf 70}, 014102 (2004);
      J. Angstmann, V. A. Dzuba, V. V. Flambaum, A. Yu. Nevsky, and 
      S. G. Karshenboim, 
      J. Phys. B: At. Mol. Phys. {\bf 39} 1937 (2006).
\bibitem{BB} K. Beloy, U. I. Safronova, and A. Derevianko,
      Phys. Rev. Lett. {\bf 97}, 040801 (2006);
      E. J. Angstmann, V. A. Dzuba, V. V. Flambaum,
      Phys. Rev. Lett. {\bf 97} 040802 (2006);
      Phys. Rev. A. {\bf 74} 023405 (2006).
\bibitem{super}
  A. Landau, E. Eliav, Y. Ishikawa, U. Kaldor,
  J. Chem. Phys. {\bf 114} 2977 (2001); 
  E. Eliav, A. Landau, Y. Ishikawa, U. Kaldor, 
  J. Phys. B: At. Mol. Phys. {\bf 35}, 1693 (2002).
\bibitem{CPM} V. A. Dzuba, V. V. Flambaum, P. G. Silvestrov, O. P. Sushkov,
   J. Phys. B:  At. Mol. Phys. {\bf 20}, 1399 (1987).
\bibitem{DSF89} V. A. Dzuba, V. V. Flambaum, O. P. Sushkov,
Phys. Lett A., {\bf 140}, 493-497 (1989);
Phys. Lett. A, {\bf 141}, 147-153 (1989);
V. A. Dzuba, V. V. Flambaum, A. Ya. Kraftmakher, O. P. Sushkov,
Phys. Lett. A, {\bf 142}, 373-377 (1989).
\bibitem{Blundell90} S. A. Blundell, W. R. Johnson, and J. Sapirstein,
Phys. Rev. Lett. {\bf 65}, 1411 (1990);
S. A. Blundell, J. Sapirstein, and W. R. Johnson,
Phys. Rev. D {\bf 45}, 1602 (1992).
\bibitem{CsPNC} V. A. Dzuba, V. V. Flambaum, J. S. M. Ginges,
Phys. Rev. D, {\bf 66}, 076013 (2002).
\bibitem{TlPNC} V. A. Dzuba, V. V. Flambaum, P. G. Silvestrov, O. P. Sushkov,
J. Phys. B: At. Mol. Phys. {\bf 20}, 3297-3311 (1987).
\bibitem{Kozlov01}
M. G. Kozlov, S. G. Porsev, and W. R. Johnson,
Phys. Rev. A 64, 052107 (2001).
\bibitem{MBPT}
Hsiang-Shun Chou and W.R. Johnson, Phys. Rev. A {\bf 56}, 2424 (1997);
S.A. Blundell, W.R. Johnson, and J. Sapirstein, Phys. Rev. A {\bf 42}, 3751 (1990).
\bibitem{CC2}
S.A. Blundell, W.R. Johnson and J. Sapirstein, Phys. Rev. A {\bf 43}, 3407 (1991);
A. Landau, E. Eliav, Y. Ishikawa, U. Kaldor,
J. Chem. Phys. {\bf 121}, 6634 (2004). 
\bibitem{CI}
K.T. Cheng, M.H. Chen, W.R. Johnson and J. Sapirstein, 
Phys. Rev. A {\bf 50}, 247 (1994);
K. T. Cheng, M. H. Chen, Rad. Phys. Chem. {\bf 75}, 1753 (2006);
P. Bogdanovich, D. Majus, T. Pakhomova, Phys. Scr. {\bf 74} 558 (2006).

\bibitem{MCDF} I. P. Grant,  Comput. Phys. Commun. {\bf 84}, 59 (1994).

\bibitem{CI+MBPT} V. A. Dzuba, V. V. Flambaum, and M. G. Kozlov,
Phys. Rev. A, {\bf 54}, 3948 (1996).

\bibitem{Ba} V. A. Dzuba and W. R. Johnson,
Phys. Rev. A, {\bf 57}, 2459 (1998).

\bibitem{hfs} V. A. Dzuba, V. V. Flambaum, M. G. Kozlov, and S. G. Porsev,
J. Exp. Theor. Phys., {\bf 87}, 885 (1998).

\bibitem{Savukov02}
I. M. Savukov, W. R. Johnson, Phys. Rev. A {\bf 65}, 042503 (2002).

\bibitem{Savukov02a}
I. M. Savukov, W. R. Johnson, H. G. Berry, Phys. Rev. A {\bf 66}, 052501 (2002).

\bibitem{Savukov03}
I. M. Savukov, J. Phys. B: At. Mol. Phys. {\bf 36}, 2001 (2003).

\bibitem{Ra} V. A. Dzuba, J. S. M. Ginges,
      Phys. Rev. A, {\bf 73} 032503 (2006);
      V. A. Dzuba and V. V. Flambaum,
      J. Phys. B: At. Mol. Opt. Phys. {\bf 40} 227 (2007). 

\bibitem{vn} V. A. Dzuba, 
      Phys. Rev. A, {\bf 71}, 032512 (2005).
\bibitem{vn1} V. A. Dzuba and V. V. Flambaum,
      Phys. Rev. A, {\bf 71}, 052509 (2005).

\bibitem{vn2} V. A. Dzuba,
      Phys. Rev. A, {\bf 71}, 062501 (2005).

\bibitem{NIST} NIST Atomic Spectra Database on Internet,
http://physics.nist.gov/cgi-bin/AtData/main\_asd.



\end{thebibliography}
\end{document}